\documentclass[10pt,conference]{IEEEtran}
\usepackage{epsfig,setspace,amsmath,epsf,amssymb,bm,theorem,cite, graphicx, epstopdf, algorithm, algpseudocode,float,color}
\usepackage{multirow}
\usepackage{authblk}
\usepackage[table,xcdraw]{xcolor}

\newtheorem{theorem}{Theorem}

\newtheorem{remark}{Remark}

\IEEEoverridecommandlockouts
\allowdisplaybreaks

\newcommand{\bx}{{\mathbf{x}}}
\newcommand{\bw}{{\mathbf{w}}}

\newcommand{\bh}{{\mathbf{h}}}

\begin{document}

\title{Improved Storage for Efficient \\ Private Information Retrieval\thanks{This work was supported by NSF Grants CNS 15-26608, CCF 17-13977 and ECCS 18-07348.}}

\author[1]{Karim Banawan}
\author[2]{Batuhan Arasli}
\author[2]{Sennur Ulukus}
\affil[1]{\normalsize Electrical Engineering Department, Faculty of Engineering, Alexandria University, Alexandria, Egypt}
\affil[2]{\normalsize Department of Electrical and Computer Engineering, University of Maryland, College Park, MD, USA}

\maketitle

\begin{abstract}
We consider the problem of private information retrieval from $N$ \emph{storage-constrained} databases. In this problem, a user wishes to retrieve a single message out of $M$ messages (of size $L$) without revealing any information about the identity of the message to individual databases. Each database stores $\mu ML$ symbols, i.e., a $\mu$ fraction of the entire library, where $\frac{1}{N} \leq \mu \leq 1$. Our goal is to characterize the optimal tradeoff curve for the storage cost (captured by $\mu$) and the normalized download cost ($D/L$). We show that the download cost can be reduced by employing a hybrid storage scheme that combines \emph{MDS coding} ideas with \emph{uncoded partial replication} ideas. When there is no coding, our scheme reduces to Attia-Kumar-Tandon storage scheme, which was initially introduced by Maddah-Ali-Niesen in the context of the caching problem, and when there is no uncoded partial replication, our scheme reduces to Banawan-Ulukus storage scheme; in general, our scheme outperforms both.
\end{abstract}

\section{Introduction}
Private information retrieval (PIR), which was introduced by Chor et al. in \cite{ChorPIR}, is a canonical problem to investigate the privacy issues that arise upon interaction with open-access databases. In classical PIR, there is a user, who needs to retrieve a message (file) out of $M$ messages from $N$ distributed \emph{content-replicating} and non-colluding databases privately, i.e., in a way that the identity of the desired message is kept secret from any individual database. A direct, yet inefficient, scheme to satisfy this privacy requirement is to download the contents of all databases. The download cost, in this case, scales linearly with $M$. Although the PIR problem was introduced in the computer science community \cite{PIRsurvey2004, cachin1999computationally}, there has been a growing interest in characterizing the fundamental limits of the problem among information theorists with notable early examples \cite{RamchandranPIR, YamamotoPIR, VardyPIR, RazanPIR, JafarPIRBlind}. Recently, Sun and Jafar have introduced the concept of PIR capacity $C$, which is the supremum of the ratio between the message size and the total download cost \cite{JafarPIR}. The optimal normalized download cost $D^*$ is the reciprocal of the PIR capacity. Sun-Jafar derived the download cost for the classical PIR model to be $D^*=1+\frac{1}{N}+\cdots+\frac{1}{N^{M-1}}$. Many interesting variants of the classical model have been studied in  \cite{JafarColluding, symmetricPIR, KarimCoded, codedsymmetric, codedcolluded,MPIRjournal, BPIRjournal, wei2017fundamental, kadhe2017private, chen2017capacity, wei2017capacity, wei2017fundamental_partial, StorageConstrainedPIR_Wei, SI_Gastpar, mirmohseni2017private, PrivateSearch, StorageConstrainedPIR, KarimAsymmetricPIR, PIR_WTC_II, noisyPIR, XSTPIR, Tian_upload, Staircase_PIR, PIR_cache_edge, Kumar_PIRarbCoded, PIR_decentralized, TamoISIT, Karim_nonreplicated, Tamo_journal}.

%PIR storage problem (motivation and other results)
In the majority of these works, the messages are replicated across the $N$ databases, such that each database stores $ML$ symbols. By leveraging this replication, the user can exploit the undesired symbols downloaded from one database as side information to recover more desired symbols from other databases. Although this simplifies the PIR scheme, it results in high storage cost. To minimize the storage cost, several directions have been explored in the literature: \cite{VardyPIR} proposes a new family of codes for storage called $k$-server PIR codes. \cite{KarimCoded} and \cite{StorageConstrainedPIR} as well store data in a non-replicated manner; \cite{KarimCoded} uses an $(N,K)$ MDS code and \cite{StorageConstrainedPIR} uses an uncoded partial replication strategy originally introduced in \cite{Caching_Maddah_Ali}. \cite{Karim_nonreplicated} and \cite{Tamo_journal} further investigate the problem of non-replicated storage via representing storage with graphs, where each database stores full messages but not the entire message set.

The works most closely related to ours are \cite{KarimCoded} and \cite{StorageConstrainedPIR}: \cite{KarimCoded} characterizes the optimal download cost of PIR from $(N,K)$ \emph{MDS-coded} databases as $D^*=1+\frac{K}{N}+\cdots+\left(\frac{K}{N}\right)^{M-1}$. In this work, each message is organized into $K$-length rows and each row is independently mapped into an $N$-length vector using an $(N,K)$ MDS code. This effectively minimizes the storage cost as each database stores $\frac{1}{K}$ of the total size of the messages. \cite{StorageConstrainedPIR} investigates a setting where each database stores a fraction $\mu$ of each message. When the storage strategy is constrained to \emph{uncoded storage}, \cite{StorageConstrainedPIR} shows that the uncoded prefetching strategy of \cite{Caching_Maddah_Ali} along with the PIR scheme of \cite{JafarPIR} is optimal and the optimal storage-download cost tradeoff is given by the convex hull of the $\left(\frac{t}{N},1+\frac{1}{t}+\cdots+\left(\frac{1}{t}\right)^{M-1}\right)$ pairs for $t=1,\dots,N$. This problem is then extended to the decentralized setting in \cite{PIR_decentralized}, where the contents are stored independently across databases according to a probability distribution, and to the heterogeneous setting in \cite{heteroPIR} where the databases have heterogeneous storage sizes. In all these works, with the exception of \cite{KarimCoded}, the fundamental limits are derived for uncoded storage strategies.

%System model
In this paper, we consider the PIR problem from storage constrained databases. The storage at the databases is constrained such that each database stores a deterministic function of the messages with a total size of $\mu ML$ symbols picked from a finite field $\mathbb{F}_q$, for some fraction $\mu$, $\frac{1}{N} \leq \mu \leq 1$. It is required to design such storage functions for facilitating the most efficient PIR scheme, i.e., we aim at jointly designing the storage strategy and the retrieval scheme such that the normalized download cost is minimized subject to a storage size constraint of $\mu ML$. The end goal is to characterize the optimal storage-normalized download cost tradeoff $D^*(\mu)$.

%Contributions
To that end, we first present a motivating example, which investigates known storage strategies for PIR, namely: uncoded storage in \cite{StorageConstrainedPIR} and direct MDS-coded storage in \cite{KarimCoded}. We show that no single storage scheme among these two schemes outperforms the other in all storage ratio regimes. Next, we restrict our coded-storage strategies to non-mixing MDS coding only, i.e., we allow message mapping to coded symbols via an MDS code that neither mixes different messages nor rows of any individual message. For an $(N,K)$ MDS code \cite{KarimCoded}, the storage ratio $\mu=\frac{1}{K}$ (as 1 coded-symbol is stored at each database from every row of the message). Therefore, the normalized download cost is $D=1+\frac{1}{N\mu}+\cdots+\left(\frac{1}{N\mu}\right)^{M-1}$, where $K=\frac{1}{\mu} \in \mathbb{Z}_+$. We aim at constructing achievable storage schemes which outperform \cite{StorageConstrainedPIR} and \cite{KarimCoded} by using a mix of MDS coding and uncoded partial replication ideas.

We propose a novel storage strategy that unifies the direct MDS coded storage in \cite{KarimCoded} with the uncoded prefetching storage in \cite{StorageConstrainedPIR}. Using this scheme, the messages are first coded row-by-row via an $(N,K)$ MDS code. The indices of the message rows are partitioned into $\binom{N}{t}$ partitions, where each row partition is stored in a group of $t$ databases. By this storage strategy, we have $\mu=\frac{t}{KN}$. We achieve a normalized download cost of $1+\frac{K}{t}+\frac{K^2}{t^2}+\cdots+\frac{K^{M-1}}{t^{M-1}}$ for all $t,K \in [N]$ and $t \geq K$. For any other point, we employ memory-sharing. For $K=1$, all corner points of the uncoded storage in \cite{StorageConstrainedPIR} can be recovered. For $t=N$, all corner points of the direct MDS-coded storage in \cite{KarimCoded} can be attained. By changing $t,K$, more corner points can be attained. At these corner points, the MDS-coded storage benchmark with rational $K$ can be achieved. We illustrate these facts by a representative example of $M=2$, $N=6$, and $\mu=\frac{5}{12}$.

\section{System Model}
\begin{figure}[t]
	\centering
	\includegraphics[width=0.45\textwidth]{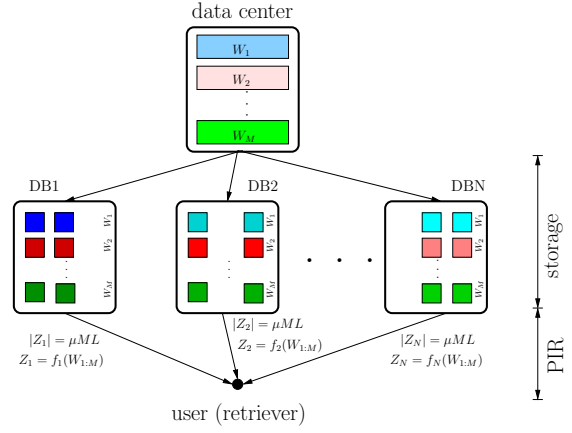}
	\caption{System model: A data center stores the message set in $N$ distributed databases; a user retrieves a message privately from the databases.}
	\label{system_model}
	\vspace{-0.5cm}
\end{figure}

Consider a storage system with $N$ databases; see Fig.~\ref{system_model}. A data center aims at storing a message set $\mathcal{W}$ in $N$ distributed databases. The message set contains $M$ i.i.d. messages of size $L$, picked uniformly from a sufficiently large field size $\mathbb{F}_q^L$,i.e.,
\begin{align}
H(W_m)&=L, \quad \text{(in $q$-ary units)} \\
H(\mathcal{W})&=H(W_1, \cdots, W_M)=ML
\end{align}
 Each database has a storage capacity of $\mu ML$, where $\frac{1}{N} \leq \mu \leq 1$. Each database stores a function of the message set $\mathcal{W}$. Specifically, the $n$th database stores $Z_n=f_n(\mathcal{W})$, such that,
\begin{align}
H(Z_n) &\leq \mu ML\\
H(Z_n|W_1, \cdots, W_M)&=0
\end{align}

In PIR, there is a user who wants to retrieve a file $W_\theta \in \mathcal{W}$, where $\theta \in [M]$, without revealing any information about $\theta$. To that end, the user submits a query $Q_n^{[\theta]}$ to the $n$th database. All the queries are independent of the messages as the user has no prior information about the message set, i.e.,
\begin{align}
I(Q_1^{[\theta]},\cdots, Q_N^{[\theta]};W_1, \cdots, W_M)=0
\end{align}

Upon receiving the query $Q_n^{[\theta]}$, the $n$th database responds with an answer string $A_n^{[\theta]}$, which is a deterministic function of the query $Q_n^{[\theta]}$ and the storage content of the database $Z_n$,
\begin{align}
H(A_n^{[\theta]}|Q_n^{[\theta]},Z_n)=0
\end{align}

To protect the privacy, the query submitted to the $n$th database to retrieve $W_\theta$ should be indistinguishable from the query submitted to retrieve $W_{\theta'}$ for all $\theta' \in [M]$, therefore,
\begin{align}\label{privacy}
(Q_n^{[\theta]},A_n^{[\theta]},\mathcal{W})\sim (Q_n^{[\theta']},A_n^{[\theta']},\mathcal{W}), \quad \theta, \theta' \in [M]
\end{align}
where $\sim$ denotes statistical equivalence.

To ensure reliability, the user should be able to reconstruct the message $W_\theta$ using the answer strings $A_{1:N}^{[\theta]}$ with arbitrarily small probability of error, hence,
\begin{align}\label{reliability}
H(W_\theta|Q_{1:N}^{[\theta]}, A_{1:N}^{[\theta]})=o(L)
\end{align}
where $\frac{o(L)}{L} \rightarrow 0$ as $L \rightarrow \infty$.

An achievable retrieval scheme is a scheme that satisfies \eqref{privacy} and \eqref{reliability} for some message length $L$ and storage functions $f_n(\cdot)$ for $n \in [N]$. We measure the efficiency of a PIR scheme with its normalized download cost $D(\mu)$
\begin{align}
D(\mu)=\frac{\sum_{n=1}^{N} H(A_n^{[\theta]})}{L}
\end{align}

In this work, we aim at jointly designing the storage system, i.e., identifying the storage functions $f_n(\cdot)$, and the retrieval scheme such that the normalized download cost is minimized. Our goal is to identify the optimal tradeoff between storage and download cost, $D^*(\mu)=\min \: D(\mu)$. We constrain ourselves to \emph{non-mixing} MDS coding based \cite{KarimCoded} storage policies.

\section{Motivating Example}\label{motivation_example}
\begin{figure}[t]
	\centering
	\includegraphics[width=0.47\textwidth]{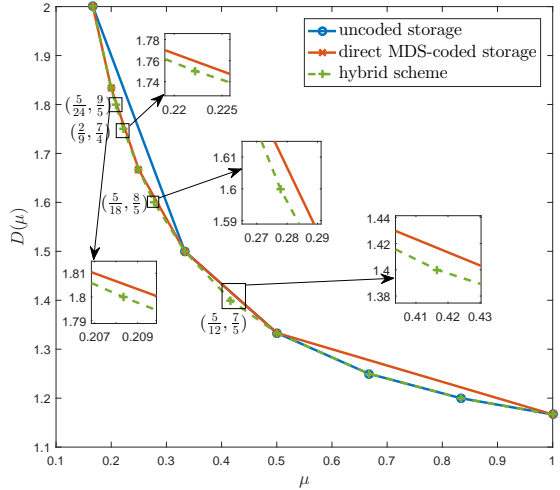}
\caption{Optimal tradeoff for uncoded storage \cite{StorageConstrainedPIR} and direct MDS coded storage \cite{KarimCoded} for $N=6$, $M=2$. The extra corner points that can be achieved only by the proposed hybrid scheme are shown explicitly on the figure and are magnified for easier visualization.}
	\label{motivation}
	\vspace{-0.5cm}
\end{figure}

In this section, we motivate our work by giving an example of a storage system with $N=6$, $M=2$. We first illustrate the uncoded storage technique in \cite{StorageConstrainedPIR}. The storage technique in \cite{StorageConstrainedPIR} is based on the uncoded prefetching scheme of \cite{Caching_Maddah_Ali}. Each message is divided into $\binom{N}{t}$ partitions such that $\mu=\frac{t}{N}$ and $t \in [N]$. For each message partition, \cite{StorageConstrainedPIR} employs the PIR scheme in \cite{JafarPIR} with $t$ databases instead of total $N$ databases. For any other storage ratio $\mu\neq \frac{t}{N}$ for some $t \in [N]$, \cite{StorageConstrainedPIR} uses memory sharing between adjacent tradeoff points that enclose $\mu$. This results in the lower convex hull of the following tradeoff points (see Fig.~\ref{motivation}, the blue curve):
\begin{align}
\left(\frac{t}{6}, 1+\frac{1}{t}\right), \quad t=1,2, \cdots, 6
\end{align}

Second, we illustrate the direct MDS coded storage technique in \cite{KarimCoded}. In \cite{KarimCoded}, each message is organized into a matrix. Each $K$-length row is mapped using an $(N,K)$ MDS code to an $N$-length vector. Each coded symbol is stored in one of the $N$ databases. Hence, $\mu=\frac{1}{K}$, for $K \in [N]$. The PIR scheme in \cite{KarimCoded} achieves the lower convex hull of the following tradeoff points (see Fig.~\ref{motivation}, the red curve):
\begin{align}
\left(\frac{1}{K}, 1+\frac{K}{6}\right), \quad K=1,2, \cdots, 6
\end{align}

From Fig.~\ref{motivation}, we note that at low storage ratios, the MDS coded storage of \cite{KarimCoded} outperforms the uncoded storage of \cite{StorageConstrainedPIR}, while the opposite is true for high storage ratios. In this paper, we propose a hybrid scheme that combines MDS coding ideas of \cite{KarimCoded} with uncoded prefetching storage ideas of \cite{StorageConstrainedPIR}. We show that our scheme outperforms both schemes for specific storage ratios. For instance, for $N=6$, $M=2$, Fig.~\ref{motivation} shows four new non-trivial corner points achieved by our scheme (see blown up sub-figures). We note that these additional four points are on the curve $D=1+\frac{1}{N\mu}+\cdots+\left(\frac{1}{N\mu}\right)^{M-1}$, so are the points already achieved by \cite{KarimCoded} and \cite{StorageConstrainedPIR}. It is unclear at this point if the entirety of this curve can be achieved by general schemes. This will likely require use of general storing strategies and corresponding PIR schemes.

\section{Main Results}
The main result of this work is a novel hybrid storage scheme tailored for the PIR problem, which outperforms the uncoded storage strategy in \cite{StorageConstrainedPIR} and the direct MDS coded storage strategy in \cite{KarimCoded}.
\begin{theorem}\label{thm1}
	For PIR from storage constrained databases, the optimal tradeoff between storage and normalized download cost $D^*(\mu)$ is upper bounded by the lower convex hull of the points:
	\begin{align}
	\left(\mu=\frac{t}{KN}, \:\:D(\mu)=D(t,K)\right), \quad t,K \in [N], t \geq K
	\end{align}
	where $D(t,K)$ is given by:
	\begin{align}\label{tradeoff}
	D(t,K)=1+\frac{K}{t}+\frac{K^2}{t^2}+\cdots+\frac{K^{M-1}}{t^{M-1}}
	\end{align}
\end{theorem}

The proof of Theorem~\ref{thm1} is given is Section~\ref{achievability}. We have the following remarks.

\begin{remark}
	Our achievable scheme achieves all the storage-download cost tradeoff points obtained by the uncoded storage systems in \cite{StorageConstrainedPIR} if we plugged in $K=1$. Similarly, our scheme achieves all tradeoff points obtained by the direct MDS coded storage systems in \cite{KarimCoded} if we plugged in $t=N$.
\end{remark}

\begin{remark}
	Using our achievable scheme, optimizing the storage problem for PIR becomes a two-dimensional optimization problem over $(t,K)$. This outperforms the storage-download cost tradeoff obtained by uncoded storage in \cite{StorageConstrainedPIR} and direct MDS coded storage in \cite{KarimCoded} as it results in a number of corner points that scales with $N^2$ \cite{Farey} in contrast with the number of corner points that scale with $N$ in \cite{KarimCoded} and \cite{StorageConstrainedPIR}, which already exist in our scheme.
\end{remark}

\begin{remark}
	We can write the tradeoff points obtained in \eqref{tradeoff} as a direct function of $\mu$ as:
	\begin{align}\label{relaxed_curve}
	D(\mu)=1+\frac{1}{N\mu}+\frac{1}{(N\mu)^2}+\cdots+\frac{1}{(N\mu)^{M-1}}
	\end{align}
	where $\mu=\frac{t}{KN}$, $t,K \in [N]$, and $t\geq K$. This expression is also achieved by the uncoded storage system in \cite{StorageConstrainedPIR} for $\mu=\frac{t}{N}$, $t \in [N]$ and by the MDS coded storage system in \cite{KarimCoded} for $\mu=\frac{1}{K}$, $K \in [N]$. It is unclear if this curve can be achieved for all $\mu \in [\frac{1}{N}, 1]$.
\end{remark}

\section{Proposed Hybrid Storage Scheme}\label{achievability}
In this section, we illustrate our achievable scheme by an example without loss of generality. We continue with the example in Section~\ref{motivation_example} where $N=6$ and $M=2$. In particular, we focus on the storage point $\mu=\frac{5}{12}=0.416$, whose download cost using the schemes in \cite{KarimCoded} and \cite{StorageConstrainedPIR}  lies on the common straight line between $\mu=\frac{1}{3}$ and $\mu=\frac{1}{2}$. Note that, for the scheme in \cite{KarimCoded} this corresponds to $K=3$ and $K=2$, and for the scheme in \cite{StorageConstrainedPIR}, this corresponds to $t=2$ and $t=3$. The download cost achieved by the schemes in \cite{KarimCoded} and \cite{StorageConstrainedPIR} on this line is $\frac{17}{12}=1.42$. Our proposed hybrid scheme achieves a strictly better download cost of $\frac{7}{5}=1.40$.

\subsection{Representative Example: $N=6$, $M=2$, $\mu=\frac{5}{12}$}
\subsubsection{Storage Phase}
Each message is organized as a $30\times 2$ matrix, with symbols picked from a sufficiently large finite field $\mathbb{F}_q$ (in order to have a feasible MDS code). Each message is coded first by a $(6,2)$ MDS code in the same manner as in \cite{KarimCoded}. Specifically, each row of each message is mapped by the $(6,2)$ MDS code into a vector of length $6$. Denote the $n$th column of the generator matrix of the $(6,2)$ MDS code by $\bh_n$, and the $j$th row of the $m$th message by $\bw_j^{[m]}$. Therefore, the MDS coded symbol corresponding to the $j$th row of the $m$th message that is intended to be stored on the $n$th database is:
\begin{align}
y_{n,j}^{[m]}=\bh_n^T \bw_j^{[m]}
\end{align}
Note that the MDS code is \emph{not mixing} the messages nor the rows of each message.

Next, we use the uncoded prefetching scheme of \cite{StorageConstrainedPIR} with $t=5$ to store the coded symbols. In this case the rows of each message is divided into $\binom{6}{5}=6$ partitions. Each partition is a non-intersecting set of row indices, whose cardinality is $5$. Let the partition $\mathcal{L}_\mathcal{S}$ be the set of row indices that should be stored in the set $\mathcal{S}$ of databases. Therefore, a possible partition assignment for our problem can be:
\begin{align}
\mathcal{L}_{\{1,2,3,4,5\}}&=\{1,7,13,19,25\}\\
\mathcal{L}_{\{1,2,3,4,6\}}&=\{2,8,14,20,26\}\\
\mathcal{L}_{\{1,2,3,5,6\}}&=\{3,9,15,21,27\}\\
\mathcal{L}_{\{1,2,4,5,6\}}&=\{4,10,16,22,28\}\\
\mathcal{L}_{\{1,3,4,5,6\}}&=\{5,11,17,23,29\}\\
\mathcal{L}_{\{2,3,4,5,6\}}&=\{6,12,18,24,30\}
\end{align}

Now, the $n$th database stores the coded symbols corresponding to the rows indexed by partitions $\mathcal{L}_\mathcal{S}$ such that $n \in \mathcal{S}$. Thus, the contents of the $n$th database can be written as:
\begin{align}
Z_n=\bigcup_{m=1}^{2} \bigcup_{\mathcal{L}_\mathcal{S}: n \in \mathcal{S}} \bigcup_{j \in \mathcal{L}_\mathcal{S}} y_{n,j}^{[m]}
\end{align}

Using this storage scheme, each database stores coded symbols from $25$ rows. Hence, $\mu=\frac{25}{2*30}=\frac{5}{12}=\frac{t}{KN}=0.42$. This point lies between $K=2$ and $K=3$ for the direct MDS coded case \cite{KarimCoded}, and between $t=2$ and $t=3$ for the uncoded prefetching case \cite{StorageConstrainedPIR}. Using the uncoded storage or the direct MDS storage, the achievable download cost is $D(\mu)=\frac{17}{12}=1.42$, which results from memory sharing.

\subsubsection{Retrieval Phase}
The user starts with permuting the row indices of each message partition independently and privately. Specifically, the user permutes $\mathcal{L}_\mathcal{S}$ randomly into two sets $\mathcal{L}_\mathcal{S}^{[1]}$ and $\mathcal{L}_\mathcal{S}^{[2]}$, where $\mathcal{L}_\mathcal{S}^{[m]}$ is the permuted set of the partition corresponding to $\mathcal{S}$ for the $m$th message. Note that these permutations are chosen independently of each other and known only at the user's side. We denote the $j$th row of the $m$th message after row permutations by $\bx_j^{[m]}$.

In round~1, the user downloads $K^2=4$ coded symbols from each database from each message. Note that the user needs to download $2$ coded symbols from the same row from $2$ different databases that contain this row, e.g., the user downloads 	$\bh_1^T \bx_1^{[1]}$, $\bh_2^T \bx_1^{[1]}$ from databases 1, 2, respectively. By MDS property, this is sufficient to decode the entire row. The main difference of this step from \cite{KarimCoded} is the fact that the user does not have the freedom to download these symbols from any 2 databases (but the databases that contain them).

Next, in round~2, the user exploits the side information generated in round~1. From round~1, the user successfully decoded $12$ rows from the undesired message. Each side information row of these is stored in $3$ databases other than the $2$ databases that originally generated this side information. Hence, the user downloads the sum of a new desired coded symbol and a side information coded symbol. Note that the user needs to download the desired coded symbol from 2 different databases in round~2 as well. The user can generate $6$ side information equations in round~2 at each database. This is due to the fact that at each database, the user downloads from 4 different rows in round~1, which the user cannot benefit from in round~2. This leaves $8$ rows to be used in round~2 as side information. Since each database stores only 10 rows out of these 12 due to the uncoded prefetching strategy, this leaves only $6$ rows to be used as side information. The complete query table is given in Table~\ref{table}.

Consequently, the user can decode the entire 30 rows of $W_1$ as the user can cancel the side information in round~1 and each desired row is decoded from $2$ different databases by the MDS property. The privacy is preserved by the random permutations of the message partitions and the identical structure of the queries for both messages. The user downloads 14 coded symbols from all databases and decode the entire $L=30*2=60$ symbols. Therefore, $D(\frac{5}{12})=\frac{14*6}{30*2}=\frac{7}{5}<\frac{17}{12}$. Hence, our proposed hybrid scheme strictly outperforms uncoded storage and direct MDS coded storage schemes.

Fig.~\ref{motivation} shows that the hybrid scheme achieves four extra corner points than direct MDS coded storage and uncoded storage. These new points are: $(\frac{5}{24}, \frac{9}{5})$, $(\frac{5}{18}, \frac{8}{5})$, $(\frac{5}{12}, \frac{7}{5})$, and $(\frac{2}{9}, \frac{7}{4})$. At all these new points, our scheme outperforms the known storage schemes and achieves the benchmark curve (\ref{relaxed_curve}) for four more values of the storage parameter $\mu$.

\begin{table*}[]
	\centering
	\caption{Query table for $M=2$, $N=6$, $\mu=\frac{5}{12}$}
	\label{table}
	\resizebox{0.88\textwidth}{!}{
	\begin{tabular}{|c|c|c|c|c|c|}
		\hline
		DB1 &DB2  &DB3  &DB4  &DB5  &DB6  \\
		\hline
		$\bh_1^T \bx_1^{[1]}$&$\bh_2^T \bx_1^{[1]}$ &$\bh_3^T \bx_2^{[1]}$& $\bh_4^T \bx_2^{[1]}$ & $\bh_5^T \bx_3^{[1]}$ & $\bh_6^T \bx_3^{[1]}$ \\
		$\bh_1^T \bx_4^{[1]}$&$\bh_2^T \bx_4^{[1]}$ &$\bh_3^T \bx_5^{[1]}$& $\bh_4^T \bx_5^{[1]}$ & $\bh_5^T \bx_6^{[1]}$ & $\bh_6^T \bx_6^{[1]}$ \\
		$\bh_1^T \bx_7^{[1]}$&$\bh_2^T \bx_7^{[1]}$ &$\bh_3^T \bx_8^{[1]}$& $\bh_4^T \bx_8^{[1]}$ & $\bh_5^T \bx_9^{[1]}$ & $\bh_6^T \bx_9^{[1]}$ \\
		$\bh_1^T \bx_{10}^{[1]}$&$\bh_2^T \bx_{10}^{[1]}$ &$\bh_3^T \bx_{11}^{[1]}$& $\bh_4^T \bx_{11}^{[1]}$ & $\bh_5^T \bx_{12}^{[1]}$ & $\bh_6^T \bx_{12}^{[1]}$ \\
		$\bh_1^T \bx_1^{[2]}$&$\bh_2^T \bx_1^{[2]}$ &$\bh_3^T \bx_2^{[2]}$& $\bh_4^T \bx_2^{[2]}$ & $\bh_5^T \bx_3^{[2]}$ & $\bh_6^T \bx_3^{[2]}$ \\
		$\bh_1^T \bx_4^{[2]}$&$\bh_2^T \bx_4^{[2]}$ &$\bh_3^T \bx_5^{[2]}$& $\bh_4^T \bx_5^{[2]}$ & $\bh_5^T \bx_6^{[2]}$ & $\bh_6^T \bx_6^{[2]}$ \\
		$\bh_1^T \bx_7^{[2]}$&$\bh_2^T \bx_7^{[2]}$ &$\bh_3^T \bx_8^{[2]}$& $\bh_4^T \bx_8^{[2]}$ & $\bh_5^T \bx_9^{[2]}$ & $\bh_6^T \bx_9^{[2]}$ \\
		$\bh_1^T \bx_{10}^{[2]}$&$\bh_2^T \bx_{10}^{[2]}$ &$\bh_3^T \bx_{11}^{[2]}$& $\bh_4^T \bx_{11}^{[2]}$ & $\bh_5^T \bx_{12}^{[2]}$ & $\bh_6^T \bx_{12}^{[2]}$ \\
		\hline
		$\bh_1^T (\bx_{13}^{[1]}+\bx_2^{[2]})$&$\bh_2^T (\bx_{13}^{[1]}+\bx_2^{[2]})$ &$\bh_3^T (\bx_{14}^{[1]}+\bx_1^{[2]})$& $\bh_4^T (\bx_{14}^{[1]}+\bx_1^{[2]})$ & $\bh_5^T (\bx_{15}^{[1]}+\bx_1^{[2]})$ & $\bh_6^T (\bx_{15}^{[1]}+\bx_2^{[2]})$ \\
		$\bh_1^T (\bx_{16}^{[1]}+\bx_3^{[2]})$&$\bh_2^T (\bx_{16}^{[1]}+\bx_3^{[2]})$ &$\bh_3^T (\bx_{17}^{[1]}+\bx_3^{[2]})$& $\bh_4^T (\bx_{17}^{[1]}+\bx_4^{[2]})$ & $\bh_5^T (\bx_{18}^{[1]}+\bx_4^{[2]})$ & $\bh_6^T (\bx_{18}^{[1]}+\bx_4^{[2]})$ \\
		$\bh_1^T (\bx_{19}^{[1]}+\bx_5^{[2]})$&$\bh_2^T (\bx_{19}^{[1]}+\bx_6^{[2]})$ &$\bh_3^T (\bx_{20}^{[1]}+\bx_6^{[2]})$& $\bh_4^T (\bx_{20}^{[1]}+\bx_6^{[2]})$ & $\bh_5^T (\bx_{21}^{[1]}+\bx_5^{[2]})$ & $\bh_6^T (\bx_{21}^{[1]}+\bx_5^{[2]})$ \\
		$\bh_1^T (\bx_{22}^{[1]}+\bx_8^{[2]})$&$\bh_2^T (\bx_{22}^{[1]}+\bx_8^{[2]})$ &$\bh_3^T (\bx_{23}^{[1]}+\bx_7^{[2]})$& $\bh_4^T (\bx_{23}^{[1]}+\bx_7^{[2]})$ & $\bh_5^T (\bx_{24}^{[1]}+\bx_7^{[2]})$ & $\bh_6^T (\bx_{24}^{[1]}+\bx_8^{[2]})$ \\
		$\bh_1^T (\bx_{25}^{[1]}+\bx_9^{[2]})$&$\bh_2^T (\bx_{25}^{[1]}+\bx_9^{[2]})$ &$\bh_3^T (\bx_{26}^{[1]}+\bx_9^{[2]})$& $\bh_4^T (\bx_{26}^{[1]}+\bx_{10}^{[2]})$ & $\bh_5^T (\bx_{27}^{[1]}+\bx_{10}^{[2]})$ & $\bh_6^T (\bx_{27}^{[1]}+\bx_{10}^{[2]})$ \\
		$\bh_1^T (\bx_{28}^{[1]}+\bx_{11}^{[2]})$&$\bh_2^T (\bx_{28}^{[1]}+\bx_{12}^{[2]})$ &$\bh_3^T (\bx_{29}^{[1]}+\bx_{12}^{[2]})$& $\bh_4^T (\bx_{29}^{[1]}+\bx_{12}^{[2]})$ & $\bh_5^T (\bx_{30}^{[1]}+\bx_{11}^{[2]})$ & $\bh_6^T (\bx_{30}^{[1]}+\bx_{11}^{[2]})$ \\
		\hline
	\end{tabular}}
	%\vspace{-0.5cm}
\end{table*}

\bibliographystyle{unsrt}
\bibliography{references_new}
\end{document}